\documentclass[journal]{IEEEtran}
\usepackage{hyperref}
\hypersetup{
colorlinks=true,
urlcolor=blue,
citecolor=blue}
\usepackage[all]{xy,xypic}
\usepackage{amsfonts,amssymb,amsmath,amsgen,amsopn,amsbsy,theorem,graphicx,epsfig}
\usepackage{eufrak,amscd,bezier,latexsym,mathrsfs,enumerate}\usepackage[utf8]{inputenc}\usepackage[english]{babel}
\usepackage[dvipsnames]{xcolor}
\usepackage[pagewise]{lineno}
\usepackage{amssymb}
\usepackage{amsmath}
\usepackage{multicol}
\usepackage{enumitem}
\usepackage{subfig}
\usepackage{multirow}
\usepackage{setspace} 
\usepackage{lipsum}
\usepackage{adjustbox}

\usepackage{optidef}

\usepackage{graphicx} 
\graphicspath{{fig/}} 
\usepackage{subfig}
\usepackage{multirow}
\usepackage{color,soul}

\usepackage{forest}
\usepackage{tikz-qtree}
\usepackage{tikz}
\usetikzlibrary{shapes.geometric}
\usetikzlibrary{arrows.meta,arrows}

\usepackage{graphicx} 
\graphicspath{{figs/}} 

\ifCLASSINFOpdf
\else
\fi
\begin{document}
%
\title{Mobile-IRS Assisted Next Generation UAV Communication Networks}
%
%
%

\author{Hazim~Shakhatreh,~\IEEEmembership{Member,~IEEE,}
        Ahmad~Sawalmeh,~\IEEEmembership{Member,~IEEE,}
        Ali H Alenezi,~\IEEEmembership{Member,~IEEE,}
        Sharief~Abdel-Razeq,~\IEEEmembership{Member,~IEEE,}
        Muhannad Almutiry,~\IEEEmembership{Member,~IEEE,}
        and~Ala Al-Fuqaha,~\IEEEmembership{Senior Member,~IEEE}
\thanks{Hazim~Shakhatreh and Sharief~Abdel-Razeq are with Department of Telecommunications Engineering, Hijjawi Faculty for Engineering Technology, Yarmouk University, Irbid, Jordan, e-mail: \texttt{(hazim.s, sharief)@yu.edu.jo}.}
\thanks{Ahmad~Sawalmeh is with Computer Science Department and Remote Sensing Unit, Northern Border University, Northern Border University, Arar, Saudi Arabia, e-mail: \texttt{ahmad.sawalmeh@nbu.edu.sa}.}
\thanks{Ali H Alenezi and Muhannad Almutiry are with Department of Electrical Engineering, Northern Border University and Remote Sensing Unit, Northern Border University,, Arar, Saudi Arabia, e-mail: \texttt{(ali.hamdan, muhannad.almutairy)@nbu.edu.sa}}
\thanks{Ala Al-Fuqaha is with  the Information and Computing Technology (ICT) Division, College of Science and Engineering, Hamad Bin Khalifa University, Doha, Qatar and Department of Computer Science, Western Michigan University, Kalamazoo, MI 49008, USA,
e-mail: \texttt{aalfuqaha@hbku.edu.qa}}
}

%
%

\markboth{}%
{Shell \MakeLowercase{\textit{et al.}}: Bare Demo of IEEEtran.cls for IEEE Journals}
%



\maketitle

\begin{abstract}
Prior research on intelligent reflection surface (IRS)-assisted unmanned aerial vehicle (UAV) communications has focused on a fixed location for the IRS or mounted on a UAV. The assumption that the IRS is located at a fixed position will prohibit mobile users from maximizing many wireless network benefits, such as data rate and coverage. Furthermore, assuming that the IRS is placed on a UAV is impractical for various reasons, including the IRS's weight and size and the speed of wind in severe weather. Unlike previous studies, this study assumes a single UAV and an IRS mounted on a mobile ground vehicle (M-IRS) to be deployed in an Internet-of-Things (IoT) 6G wireless network to maximize the average data rate. Such a methodology for providing wireless coverage using an M-IRS assisted UAV system is expected in smart cities. In this paper, we formulate an optimization problem to find an efficient trajectory for the UAV, an efficient path for the M-IRS, and users’ power allocation coefficients that maximize the average data rate for mobile ground users. Due to its intractability, we propose efficient techniques that can help in finding the solution to the optimization problem. First, we show that our dynamic power allocation technique outperforms the fixed power allocation technique in terms of network average sum rate. Then we employ the individual movement model (Random Waypoint Model) in order to represent the users’ movements inside the coverage area. Finally, we propose an efficient approach using a Genetic Algorithm (GA) for finding an efficient trajectory for the UAV, and an efficient path for the M-IRS to provide wireless connectivity for mobile users during their movement. We demonstrate through simulations that our methodology can enhance the average data rate by 15\% on average compared with the static IRS and by 25\% on average compared without the IRS system.

\end{abstract}


%
\IEEEpeerreviewmaketitle

\section{Introduction}
%
%
%
%
\IEEEPARstart{I}{ntelligent} reflection surface (IRS) is considered an emerging technology for future wireless networks. The IRS technology's main advantages in terms of boosting the performance of wireless communication are as follows \cite{yan2021intelligent}: 1) IRS is extremely compatible with various wireless physical layer technologies that are currently being developed since it focuses on signal propagation over a wireless medium, while the other approaches are mostly used in transceivers; 2) IRS is made up of a massive number of small reflecting elements arranged on a planar surface, and it has a minimal hardware complexity and a lot of flexibility in terms of deployment; 3) The reflecting elements of the IRS are passive, requiring no active radio-frequency (RF) chains that consume a lot of power, and control circuits for the IRS are likewise ultra-low-power electronic circuits \cite{wu2019towards}. As a result, IRS is an energy-saving device that could be powered by wireless energy harvesting. On the other hand, one of the essential uses for UAVs is wireless communications, which is projected to play a critical part in future wireless networks \cite{mkiramweni2018game}. UAVs with wireless transceivers can be utilized as relays for data transmission. It can be used as an aerial base station to provide services to places without access to the internet. UAVs can also be utilized to gather and deliver information between ground stations and during mining surveying and exploration missions. 

Non-orthogonal multiple access (NOMA) has been recognized as a promising multiple access candidate for the six-generation (6G) networks \cite{liu2021application}. NOMA outperforms traditional orthogonal multiple access (OMA) in various ways, including \cite{islam2017noma}: 1) It delivers higher spectrum efficiency by simultaneously serving numerous users with the same frequency resource and minimizing interference through successive interference cancellation (SIC); 2) It increases the number of users serviced at the same time, allowing for massive connectivity; 3) Due to the nature of simultaneous transmission, a user does not need to go through a predefined time slot to transmit their data, resulting in shorter latency; 4) Flexible power control between strong and weak users allows NOMA to preserve user fairness and diverse quality of service \cite{wei2016survey}; in particular, because more power is allocated to a weak user, NOMA offers higher cell-edge throughput and so improves the cell-edge user experience.

Previous studies on IRS-Assisted UAV Communications typically consider a fixed placement for the IRS or to be mounted on a UAV. However, the assumption of placing the IRS at a fixed position will prevent the mobile users from enhancing many benefits from the wireless network, such as data rate, coverage, etc. Moreover, the assumption of placing the IRS on a UAV is not practical for many reasons including: the heavy IRS weight and its large size, and the speed of wind in bad weather can effects the stability of the UAV in the sky. The authors in \cite{cai2020resource} utilize an IRS to assist a UAV in serving multiple ground users. They aim to reduce the system's average overall power consumption by optimizing the resource allocation approach, the velocity and trajectory of the UAV, and IRS's phase control. The IRS's restricted energy budget and the individual minimum data rate requirement are combined as constraints in a non-convex optimization problem. To find an effective suboptimal solution, they suggest an alternating optimization algorithm. In \cite{wei2020sum}, the authors study the use of IRS in UAV-based OFDMA communication systems, which takes advantage of both the IRS's considerable beamforming gain and the UAV's high mobility to improve the system sum rate. The proposed system's joint design of the UAV's trajectory, IRS scheduling, and communication resource allocation is formulated as a non-convex optimization problem to optimize the system sum-rate while accounting for each user's heterogeneous quality-of-service (QoS) requirements. To address the lower bound optimization problem, an alternating optimization algorithm is created, and its results are compared to the benchmark results produced by addressing the upper bound problem. The authors in \cite{jiao2020joint} consider an IRS-based UAV-assisted multiple-input single-output NOMA downlink network. Their goal is to increase the strong user's rate while ensuring the weak user's target rate, which is determined by the optimal UAV horizontal position. They begin by optimizing the IRS-UAV location. Next, the authors suggest an iterative approach for optimizing IRS transmit beamforming and phase shift. The closed-form formulas of the best beamforming vectors are derived. They next propose two strategies for obtaining the best IRS phase-shifting based on the obtained beamforming. In \cite{mahmoud2021intelligent}, the authors investigate the performance of mounted IRS-assisted UAV communication. An end-to-end path loss model based on the practical elevation angle-dependent path loss model was used for base station-UAV and UAV-ground user equipment links. The asymptotic SNR analysis was compared to the theoretical bounds on the average SNR. The authors in \cite{shafique2020optimization}, studied the end-to-end performance of IRS-UAV relay system in terms of SNR outage probability, ergodic capacity, and energy efficiency in three types of modes: IRS mode, UAV-IRS mode, and UAV mode. They optimized the number of IRS elements and UAV height for the IRS mode, and they optimized the height for the UAV in the UAV mode. They find that the optimum height differed depending on the transmission mode. They have also presented an analytical criterion for determining the most energy-efficient height and mode. In \cite{mamaghani2021aerial}, the authors propose a THz-band IRS communication system mounted on a UAV for distributing data from an access point to multiple ground users in IoT networks. They formulate an optimization problem in terms of the minimum average energy efficiency to improve communication while minimizing network power usage. They present a computationally efficient approach that iteratively solves a sequence of approximated convex sub-problems using a block coordinated successive convex approximation. With complexity and convergence analysis, they suggest a low-complex overall approach for improving system performance.

In this paper we utilize a single UAV, and an IRS mounted on a mobile ground vehicle (M-IRS) to maximize the average data rate in an IoT-6G wireless network. The aim of the formulated problem is to find an efficient trajectory for the UAV, an efficient path for the M-IRS, and users' power allocation coefficients that maximize the average data rate for mobile ground users in a 6G wireless network. As far as we know, this is the first study that proposes utilizing an M-IRS and a UAV to maximize the average data rate in an IoT-6G wireless network. The following are the primary contributions of this paper:
\begin{itemize}
 
  \item Realistic path models are utilized for wireless connections. The ground user receives a signal from the UAV in line-of-sight (LoS) and non-line-of-sight (NLoS) paths. Moreover, it is assumed that the ground user receives signals from NLoS paths reflected by M-IRS elements due to the environmental obstacles.
  
  
  \item The optimization problem is formulated to find an efficient trajectory for the UAV, an efficient path for the M-IRS, and users' power allocation coefficients that maximize the average data rate for mobile ground users in an IoT-6G wireless network.
  
  \item  We show that the proposed dynamic power allocation technique outperforms the fixed power allocation technique in terms of network average sum rate.
  
  \item The individual movement model (Random Waypoint Model) is employed to represent the users’ movements inside the coverage area. In this model, the ground-moving users can randomly move and change their location without any constraints. Moreover, the users’ velocity, movement direction, and next location are selected randomly and separated from other users in the group.
  
  \item An efficient approach is proposed using a Genetic Algorithm (GA) for finding an efficient trajectory for the UAV and an efficient path for the M-IRS to provide wireless connectivity for mobile users during their movement.
\end{itemize}

The rest of the paper is organized as follows. The system model, including the problem formulation, is presented in Section~\ref{sysmodel}. Section~\ref{sec3} introduces the dynamic power allocation strategy. In Section~\ref{sec4}, we present the proposed mobility model for users during their movement.  In Section~\ref{sec5}, we propose the Genetic Algorithm to determine an efficient UAV trajectory, and an efficient path for the M-IRS. In Section~\ref{sec6}, we provide the results of our experiments. Section~\ref{lessons} presents the lessons learned from this research.  Finally, Section~\ref{sec7} concludes the paper. 

\section{System Model and Problem Formulation }
\label{sysmodel}
Consider a geographical area served by an M-IRS and a UAV, as shown in Figure 1. Let the 2D location of the M-IRS at time $t\in T$ is denoted by $(x_{R}^t, y_{R}^t)$ and let the 3D location of the UAV at time $t\in T$ is denoted by $(x_{D}^t, y_{D}^t, z_D^t)$. The UAV's movement is restricted to specified heights based on legal constraints, whereas the roads in the given area constrain the M-IRS's mobility. We assume that all users are located inside the geographical area and use $(x_{i}^t, y_{i}^t)$ to denote the location of user $i\in I$ at time $t\in T$. Note that the users can change their locations every time slot $t\in T$. The M-IRS and the UAV cooperate together to maximize the sum-rate for all users in this network. Hence, we need to find the efficient path for the M-IRS and the efficient trajectory for the UAV.

\begin{figure*}[!h]
\centering
\includegraphics[width=6.5in]{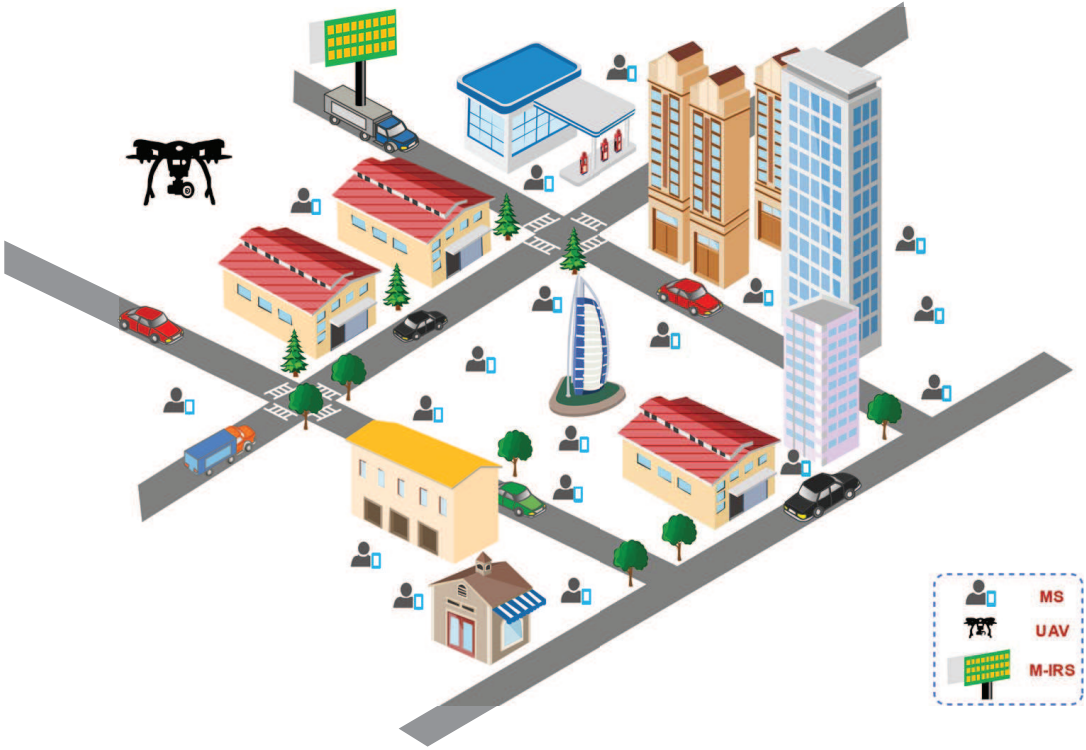}
\caption{M-IRS UAV NOMA Network.}
\label{fig:model}
\end{figure*}

The M-IRS and the UAV have single antennas. A downlink scenario is being considered in which the NOMA technique is utilized to transmit information and ensure wireless coverage for ground users. The M-IRS is made up of N elements, each of which may be dynamically programmed and changed to regulate the reflected signal's direction and phase \cite{ma2020enhancing}. Once the ground user's location is known, the controller may calculate the direction and phase shift of each M-IRS element's reflected signal so that these signals can be coherently combined at the ground user. Both the direction and phase change are assumed to be perfect and controlled continuously. The ground user receives signals from the UAV's LoS and NLoS paths. Also, it is assumed that the ground user receives signals from NLoS paths reflected by M-IRS elements due to the obstacles of the environment.

The received signal at the ground user in such a network can be written as follows using the NOMA-IRS concept:
\begin{equation}
r_i^t=\sum_{i\in I}\sqrt{\alpha_i^tP}h_{i}^ts_i^t+\displaystyle \sum_{n\in N}h_{n}^ts_i^t+v,
\end{equation}
where the symbol that the UAV transmits is $s_i^t$, $P$ is the UAV's maximum transmission power, and $\alpha_i^t$ represents the $i$th user's power allocation coefficient.  At the ground user, $v$ signifies additive white Gaussian noise (AWGN) with a variance of $\sigma^2$ and a mean of zero, $h_i^t$ is the gain of the channel between the UAV and the ground user, and $h_{n}^t$ is the channel gain between the element of the M-IRS and the ground user. The received signals can be constructively combined to create increased received signal power with the ability to alter the phase of the reflected signal using M-IRS. Based on feedback from ground users, the UAV and the M-IRS are assumed to be totally aware of the channel gain. Furthermore, the ground users perform a successive interference cancellation (SIC) operation to detect the UAV's symbols. For perfect SIC, the channel gains can be sorted as follows, without loss any generality, based on channel gains information:
$|h_1^t|^2\leq|h_2^t|^2 \leq...\leq|h_{|I|}^t|^2$. Hence, $\alpha_1^t\geq\alpha_2^t\geq...\geq\alpha_{|I|}^t$.

The average pathloss between the UAV and the $i$th user is given by
\cite{gapeyenko2018effects}:
\begin{equation}
L_i^t=P_{L,i}^tL_{L,i}^t+P_{N,i}^tL_{N,i}^t,
\end{equation}
where $L_{L,i}^t$ and $L_{N,i}^t$ are the pathlosses for LOS and NLOS paths, respectively. They are given by the  following formulas:
\begin{equation}
L_{L,i}^t=a_L+10b_L\log_{10}d_i^t,
\end{equation}
\begin{equation}
L_{N,i}^t=a_N+10b_N\log_{10}d_i^t,
\end{equation}
in which $a_L,b_L,a_N,b_N$ are the parameters of the
LOS and NLOS pathlosses, and \[ d_i^t=\sqrt{(x_{D}^t-x^t_i)^2+(y_{D}^t-y^t_i)^2+(z_{D}^t)^2}\] is the distance between the $i$th ground user and the UAV. Furthermore, for the $i$th user, $P_{L,i}^t$ can be modeled as the chance of human body blockage and it is given by \cite{gapeyenko2016analysis}:
 \begin{equation}
P_{L,i}^t=\exp\Bigg(-\lambda g_B\dfrac{q_i^th_B^t}{z_D^t}\Bigg) ,
 \end{equation}
where $q_i^t$ is 2D distance between the projection of the UAV onto the ground and the $i$th ground user at time $t\in T$, $z_D^t$ is the UAV's altitude, $\lambda$ is the human blocker's density, $g_B$ is the human blocker's diameter, $h_B^t$ is the human blocker's height. 

The path between the M-IRS and the ground user is assumed to be NLOS and the pathloss model is given by \cite{akdeniz2014millimeter}:
\begin{equation}
L_{N,i}^t=a_N+10b_Nlog_{10}d_i^t.
\end{equation}

According to the power allocation coefficients, the ground user decodes the signals of other users. Hence, at the $i$th user ($1\leq i\leq I-1$), the received signal-to-interference-plus-noise-ratio ($\gamma_i^t$) can be written as \cite{kizilirmak2016non}:
\begin{equation}
\gamma_i^t=\dfrac{\alpha_i^t|h_i^t|^2+\beta_i^t|h_n^t|^2}{\sum_{j=i+1}^{|I|}\alpha_j^t|h_j^t|^2+1/\rho},
\end{equation}
where $\rho=P/\sigma^2$ is the transmit SNR. Moreover, the term $\sum_{j=i+1}^{|I|}\alpha_j^t|h_j^t|^2$ represents the interuser interference after SIC, and $\beta_i^t$ power reflection coefficient from M-IRS. Since the NOMA method divided the $I$ users into $I/2$ pairs, each pair will occupy one sub-band channel. The total UAV transmits power $P_{\text{UAV}}$ is divided equally between the $I/2$ pairs. Then, each sub-band will distribute the assigned power between the pair users according to the fractional transmit power allocation (FTPA) method, such that each sub-band will have a power fraction as follows~\cite{benjebbovu2013system}:

\begin{equation}
\alpha_i = \dfrac{(|h_i^t|^2/\sigma_i^2)^\beta}{\sum_{j=1,2} (|h_j^t|^2/\sigma_j^2)^\beta},
\end{equation}

where $\beta$ is the decay coefficient, it has a value between 0 and 1. If its value is zero, it means that the power is divided equally among the users, and if its value increases, the user with weaker channel $|h_i^t|^2$ will have a greater power fraction. Therefore, the value of $\beta$ is optimized via simulation until we find the value maximizes system performance.

As a result, the total network's sum-rate during the coverage process will be as follows:
\begin{equation}
R=\displaystyle \sum_{i\in I}\sum_{t\in T} \log_2\big[1+\gamma_i^t \big]
\end{equation}
To find the maximum sum-rate in this system during the coverage process, the
trajectory of the UAV, the path of the M-IRS, and the users' power allocation coefficients
should be jointly optimized. The formulation of the optimization problem can be written as:

\begin{equation}
\begin{split}
\max_{(x_{D}^t, y_{D}^t, z_D^t),(x_{R}^t, y_{R}^t),a_i^t } \displaystyle \sum_{i\in I}\sum_{t\in T} \log_2\big[1+\gamma_i^t\big]~~~~~~~~~~~~~~~~~~~~~~~\\
subject ~to~~~~~~~~~~~~~~~~~~~~~~~~~~~~~~~~~~~~~~~~~~~~~~~~~~~~\\
\gamma_i^t\geq \gamma_{\text{th}},~~~~~~~~~~~~~~~~~ \forall i \in I, \forall t \in T,\\
(x_{D}^t, y_{D}^t, z_D^t)\in {\cal R}^3,~~~~~~~~~~~~~~~~~~~~ \forall t \in T,\\
(x_{R}^t, y_{R}^t)\in {\cal R}^2,~~~~~~~~~~~~~~~~~~~~~~~~ \forall t \in T,\\
\sum_{i\in I}a_i^t\leq 1,~~~~~~~~~~~~~~~~~~~~~~~~~ \forall t \in T,\\
0\leq \theta_{i,n}^t\leq 2\pi,~~~~\forall i \in I, \forall t \in T, \forall n \in N,\\
\end{split}	
\end{equation}
The first constraint set ensures that the ground user's data rate is always greater than or equal to the threshold value. The second and third constraint sets show the permissible values for 3D and 2D locations of the UAV and M-IRS, respectively. The fourth constraint set guarantees that the overall transmitted power of the UAV is not greater than its maximum power. The last constraint set is the angle constraint of each element on M-IRS. 

Due to many unknowns and nonlinear constraints, solving the optimization problem is challenging. Furthermore, the optimization problem must be performed over an infinite number of feasible UAV and M-IRS positions in a continuous space. Therefore, there is a need for efficient algorithms that can help in finding the solution to the optimization problem.

\section{NOMA Pairing}
\label{sec3}

In NOMA networks, users are often divided into two categories, weak and strong users. Weak users are the users with low channel gain, such users usually the ones that are located faraway from the BS, whereas strong users are quite the opposite, i.e.,  the users that have high channel gain and usually located near the BS \cite{shahab2016user,abdel2021pso}. Furthermore, grouping users is a useful strategy for reducing system complexity \cite{ding2015impact}. Typically, each group consists of two users with varying channel gains. This method is frequently used in previous research studied the NOMA networks \cite{gao2017theoretical,ding2016general,wei2017optimal,islam2015non,liang2019non,tang2017performance,sedaghat2018user}. This can be justified as follows: it's will known that the strong user in each group must execute SIC. Hence, having more than two users per group introduces significant computing complexity, which hurts the implementation of NOMA. The other reason is that the SIC decoding of the signals is performed sequentially, which means grouping more users into each group would increase latency and may introduce SIC error propagation. Such problems are not accepted in 5G-and-Beyond wireless networks. \cite{zeng2018energy,al2022energy}\footnote{It's important noting that communications across user groups are maintained orthogonal by employing frequency or time division, i.e., FDM or TDM. \cite{abdel2019superposition}}. However, the fundamental problem with NOMA networks is determining which users belong to which groups. As a result, several researchers have developed various strategies to maximize the total network rate while minimizing SIC errors. Such pairs can be discovered using a brute-force search of all potential sets of accessible pairs or by simply pairing users with significantly different channel gains. In the former case, this would add additional calculation load to the system, but the latter is far easier to implement and has been used in the majority of current work on this issue \cite{al2016user,chen2017user,zhang2018performance}. Even though the rest of this paper assumes two users per group, it can be generalized for any number of users per group.

\section{Mobility Models}
\label{sec4}

The mobility models are developed to define the motion and mobility for users during their movement. Specifically, it represents the changes in the user's location. Mobility models are divided into two classes; individual movement models  \cite{camp2002survey} and group movement models \cite{wang2002group}.
	
	In this work, the individual movement model, Random Waypoint  \cite{broch1998performance} is employed to represent the users'  movement inside the coverage area.
	In this model, the ground-moving users can randomly move and change their location without any constraints. Moreover, the users' velocity, movement direction, and next location are selected randomly and separated from other users in the group \cite{broch1998performance}. 
More specifically, in the coverage region, the ground receivers are distributed randomly. Then, the ground receivers motion can be expressed as following steps: 
\begin{enumerate}
\item The ground receivers select their next locations randomly, this refers to a waypoint. 
\item The ground receivers speed are randomly picked from the predefined speed period of [$S_{\min},  S_{\max}$].
\item Every ground receiver advances towards its picked next location.
\item When the ground receiver arrives at his next location, he paused for a constant period of time. This motion represents the ground receiver movement pattern that remains at a placement for a specific time before it proceeds to a next location.
\item After that, the ground receiver selects another next location. Then, the steps from  1 to 4 will be repeated until the users arrive at stationary distribution \cite{aschenbruck2008survey}.
\end{enumerate}
\section{UAV and M-IRS Trajectory Algorithm }
\label{sec5}

In this section, an efficient approach using a Genetic Algorithm (GA) is proposed for finding an efficient trajectory for both UAV and IRS to provide wireless connectivity for mobile users during their movement such that the average data rate for mobile ground users in an IoT-6G wireless network is maximized.

\subsection{Genetic Algorithm (GA)}
	A Genetic Algorithm is a meat-heuristic approach developed from Darwin natural evolution and selection theory. GA could be utilized to discover near-optimal solutions that solve NP-hard and non-convex optimization problems \cite{goldberg2006genetic,sivanandam2008genetic}.
	In this work, GA is employed for finding near-optimal trajectories for UAV and M-IRS that maximize the data rate. Specifically, the optimization problem is formulated to obtain an efficient trajectory for the UAV, an efficient path for the M-IRS, and users’ power allocation coefficients that maximize the average data rate for mobile ground users in an IoT-6G wireless network.
	
    Moreover, the GA phases are discussed \cite{goldberg2006genetic}, and how the GA algorithm is employed to find a near-optimal solution to the optimization problem.
	
	The GA consists of five steps: initial population, fitness function, selection, crossover, and mutation. The purpose of each step is demonstrated below: 
	
	\begin{enumerate}[label=\alph*.]
	    \item \textbf{Initial population:} 	The Random generation process starts with a set of individuals which is called initial population $N_{pop}$. Each individual expresses a reasonable solution to the proposed optimization problem. A group of parameters describes the individual, such as a combination of numerals, letters, or symbols called a chromosome. 
\item \textbf{Fitness function:} This function assesses each individual in the population. Specifically, it determines how good the candidate solution is.
	The fitness function continually computes the fitness value for individuals. In every iteration, GA  assigns a fitness score to every individual. After that, GA sets the fitness values according to their scores.
    Finally, the selection stage utilized these scores to produce the next generation. In this stage, the higher scores individuals have more chance to be chosen. 
    
\item \textbf{Selection:} At this point, a set of the fittest individuals called parents is chosen for the next generation; here, the offspring inherit their parent's genes. 
    In this work, tournament selection strategy is used to select the fittest individuals from the current generation. The selected candidates are then passed on to the next generation.  
	
    \item \textbf{Crossover:}  This is a fundamental GA operation. For each mated pair's parents, a crossover point is randomly chosen from the genes inside the chromosome.  Offspring are generated by exchanging the parent's genes until reached the crossover point. 
	After that, the generated offspring produce the next generation then, it will be added to the existing population.
    
    \item \textbf{Mutation:} The mutation stage ensues the diversity for the generated population. Moreover, it will be used to prevent prematurely for GA algorithm. Specifically, the solution of the optimization problem converges to the local optimal point. This paper utilizes a bit flip mutation approach.  
	\end{enumerate}

\section{Simulation Results }
\label{sec6}

In this work, a single UAV is deployed as an aerial base-station; moreover, a mobile IRS is considered to provide wireless connectivity for ground terminals inside the coverage region ${\cal R}$. This section shows the results of the proposed approach to find efficient trajectories for M-IRS and UAV that consider the mobility of the ground terminals based on the random waypoint movement model.
Specifically, the proposed method is employed to find an efficient trajectory for the UAV and an efficient path for the M-IRS that maximizes the average data rate for mobile ground terminals in a 6G wireless network. 

\begin{figure}[!h]
		\centering
		\includegraphics[scale=0.35]{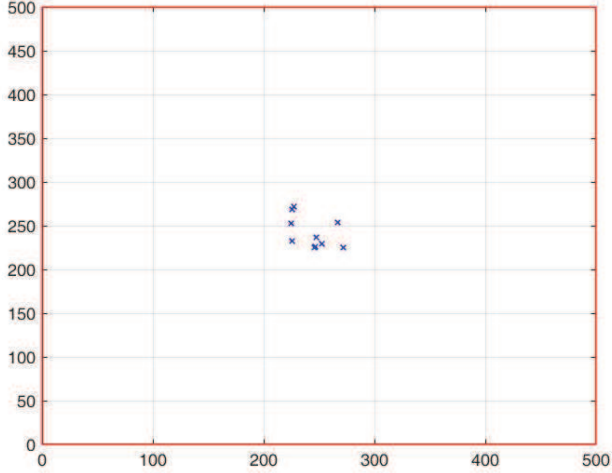}
		\caption{Users Distribution inside the coverage region ${\cal R}$ at $t_1$.  The axes' dimensions are in meters.}
		\label{usrs_distr}
	\end{figure}

	Consider a square coverage area denoted as ${\cal R}$, where the ground terminals are initially distributed inside a small subregion $50\times50$ $meters$ as illustrated in Figure~\ref{usrs_distr}.

\begin{figure*}[!h]
  \centering
   \subfloat[Subregion at $t_2$.]{\label{figur:1}\includegraphics[width=.45\textwidth]{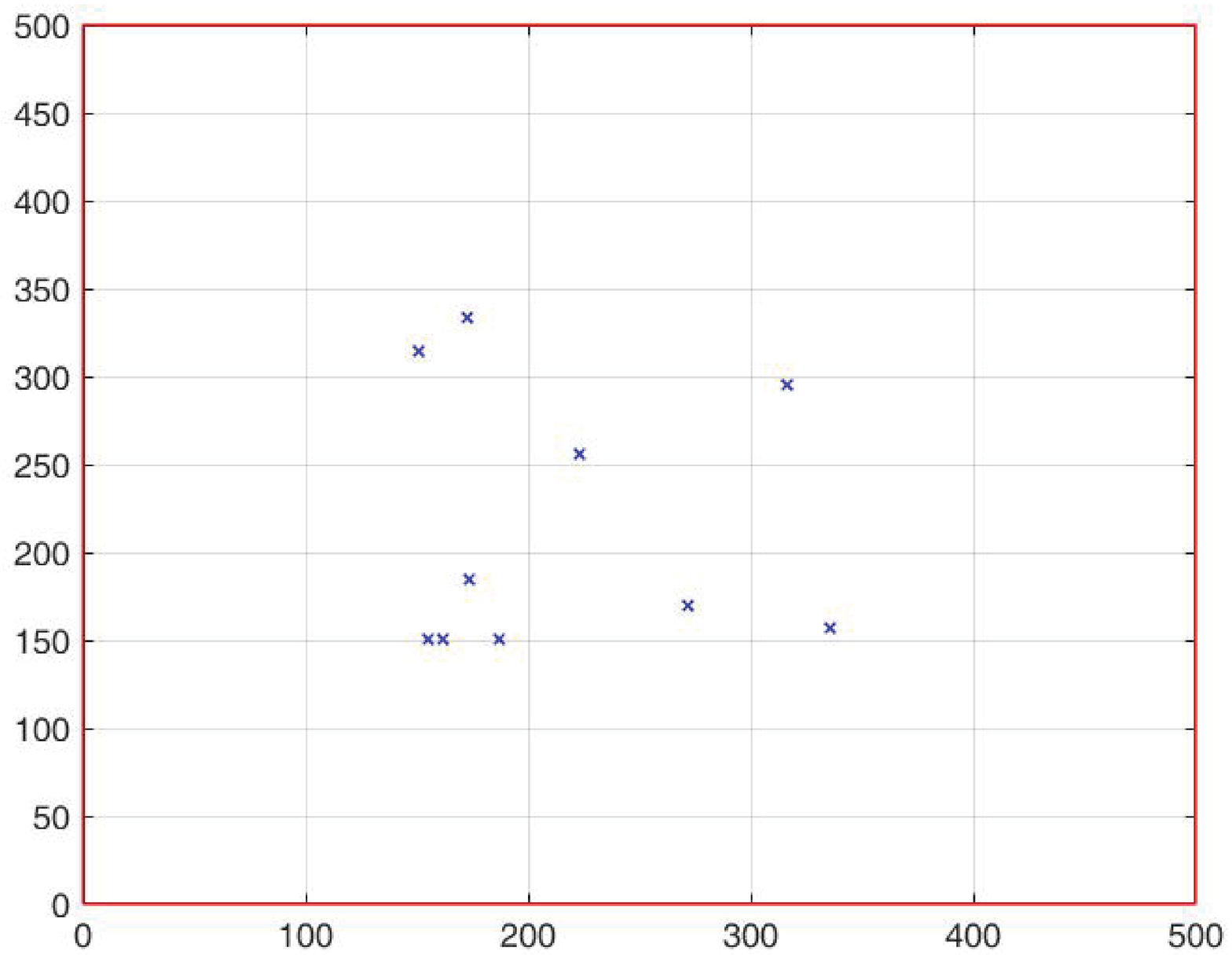}}
  \subfloat[Subregion at $t_3$.]{\label{figur:2}\includegraphics[width=.45\textwidth]{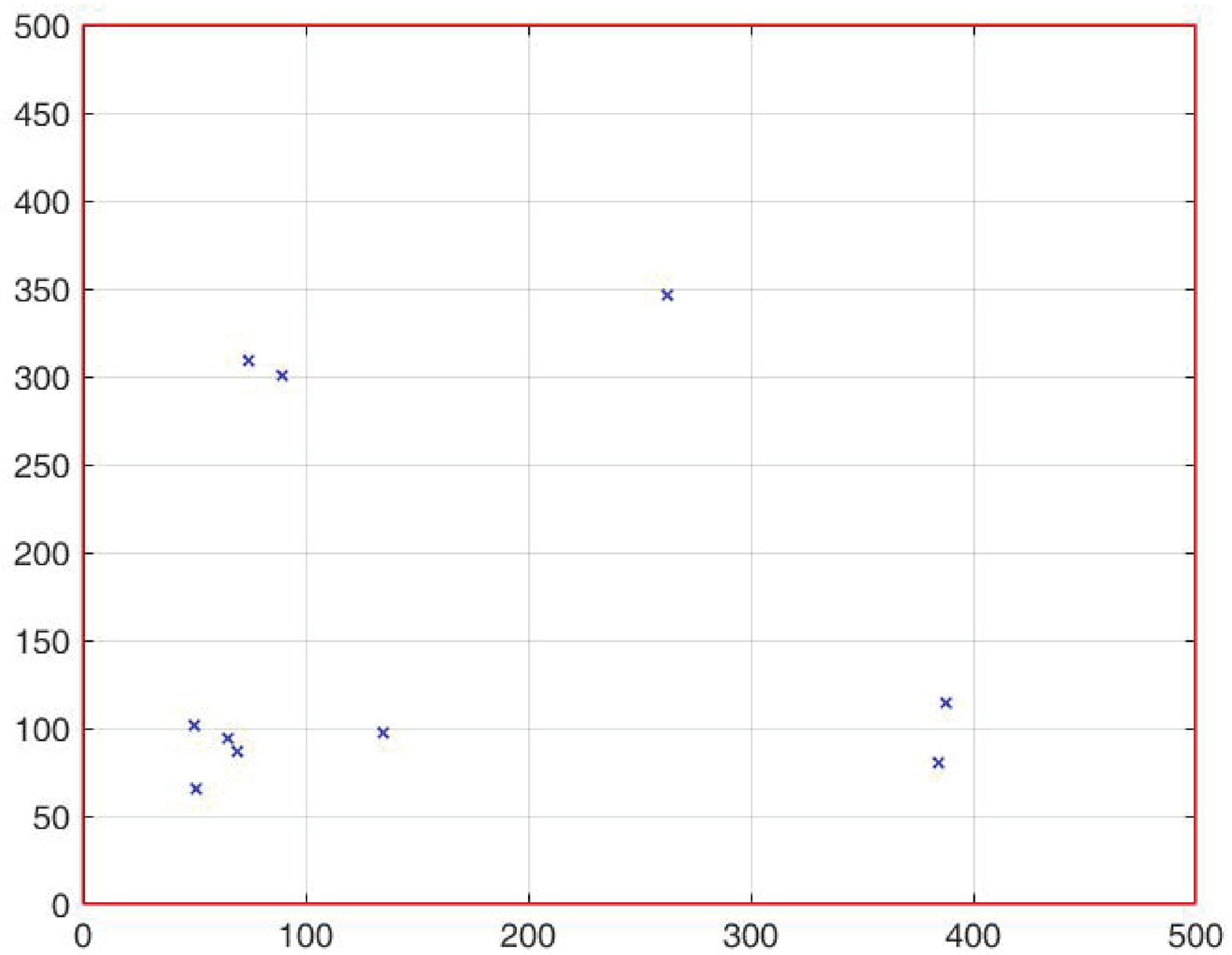}}
  \\
  \subfloat[Subregion at $t_4$.]{\label{figur:3}\includegraphics[width=.45\textwidth]{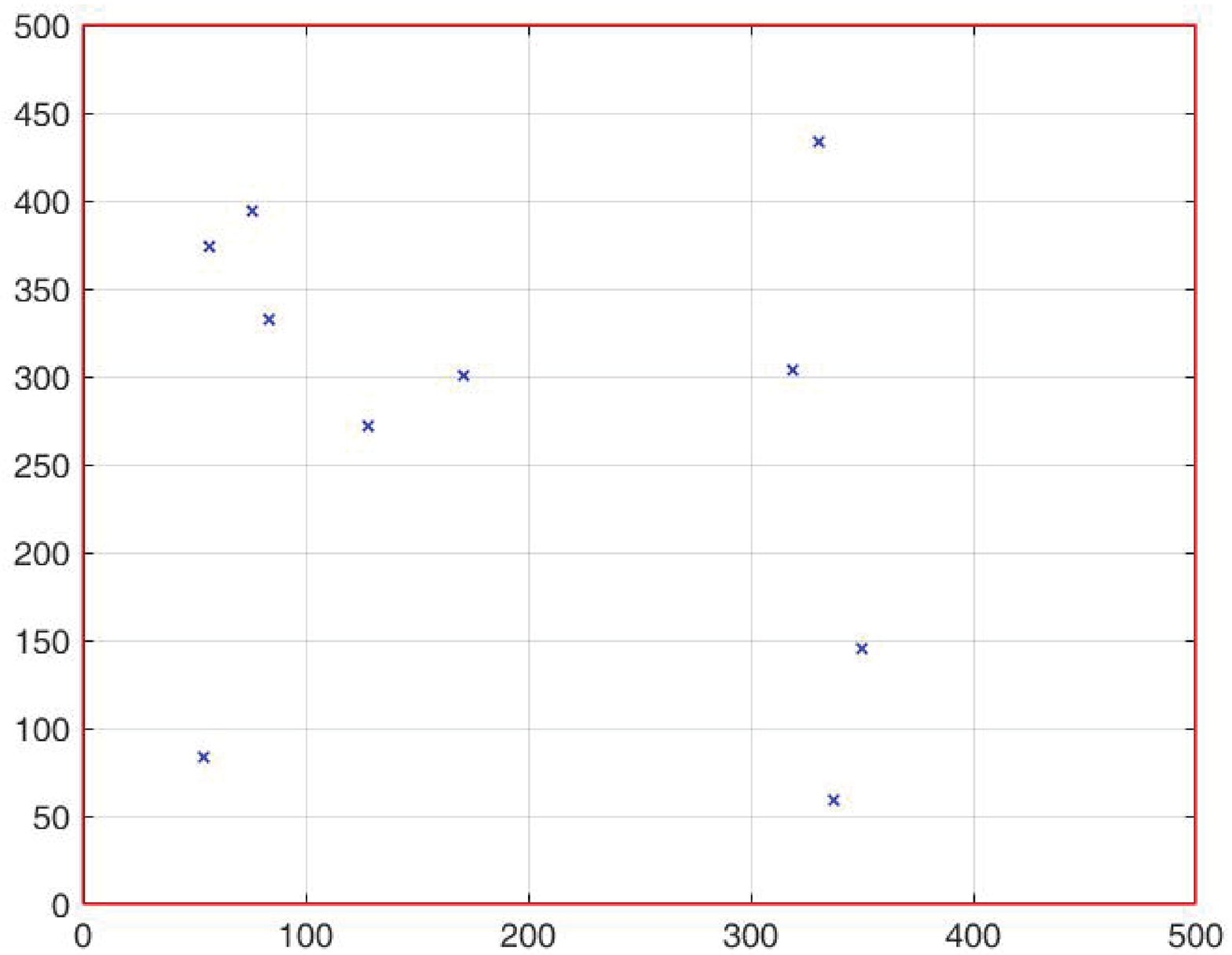}}
  \subfloat[Subregion at $t_5$.]{\label{figur:4}\includegraphics[width=.45\textwidth]{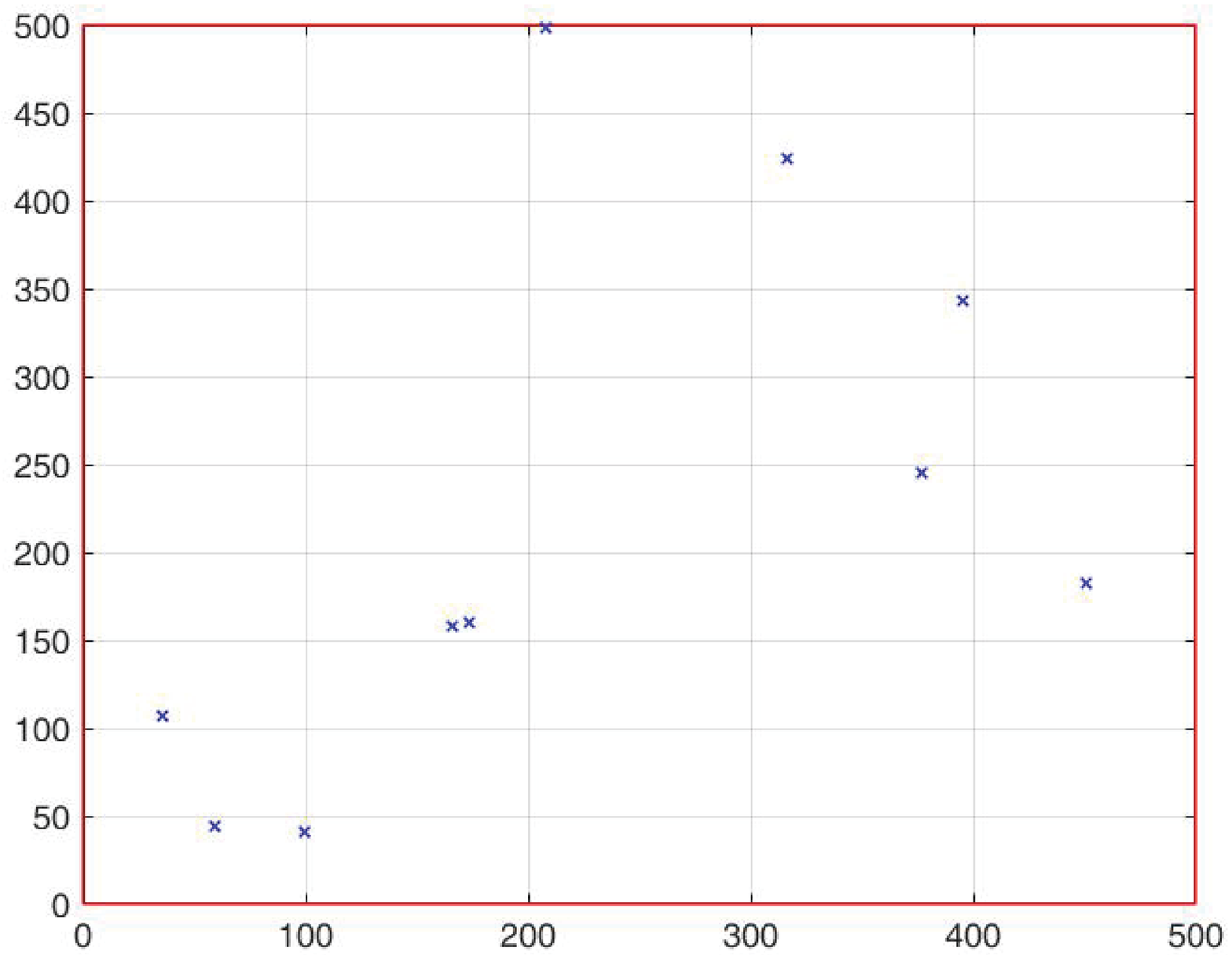}}
 \caption{Users distribution and trajectory within a Subregion  from $t_2$ to $t_5$ $\in$ $T$. This Figure represents the users' movement from time $t_2$ to $t_5$ based on the random waypoint mobility model,  where the time period and the users' locations are updated every 5 minutes. }
  \label{figur2-2}
\end{figure*}

	The ground terminals are moving based on the random waypoint mobility model; their movement is represented in Figure~\ref{figur2-2}, where the time period is updated every 5 minutes.

More specifically, in this work, 
there are five times slots, and the users are moving based on the random waypoint model inside the coverage region. Figure~\ref{usrs_distr} presents the users' distribution at the first time slot $t1$. Moreover, Figure~\ref{figur2-2} shows the users distribution and movement from T2 as in Figure~\ref{figur:1} to T5 Figure~\ref{figur:4}. 
The UAV and the M-IRS locations are updated at the end of every time slot. 
In this work, each user receives a line of sight signal from the UAV and a reflected signal from the M-IRS. The M-IRS contains $K \times N$ elements; each user will benefit from $N$ M-IRS elements to strengthen their received signal. The NOMA technology is utilized for this work; therefore, we assume that only two users share the same frequency band. Furthermore, we assume the SIC will perfectly remove the interference; thus, there is no error propagation. The simulation parameters are presented in the following Table~\ref{sim_par}. 
 

	\begin{table*}[!h]
		\renewcommand{\arraystretch}{1.2}
		\label{table1}
		\centering
		\caption{{\MakeUppercase{The Simulation Parameters}}}
		\begin{adjustbox}{width=1.0\textwidth}
			\begin{tabular}{|l|l|l||l|l|l|}
				\hline
				\multicolumn{3}{|c||}{Simulation  Parameters}&\multicolumn{3}{|c|}{System and Algorithms Parameters} \\
				\hline
				Subregion $({\cal R})$ dimensions &$(x_{1}, y_{1})$,$(x_{2}, y_{2})$&$(0,0)$, (500m, 500m) &Carrier frequency& $f_c$  & 28 GHz \\
				\hline
				Number of ground terminals&$M$&$10$ &Noise power &$Np$ & -80 dBm \\
	
				\hline
				 
				 Max. UAV transmission power& $P_{t_{\text{UAV}_{max}}}$& $36~$ dBm &GA Population size &$N\_pop$ &50\\
				
				\hline
				Signal to Noise Theshold&SNR&20 dB & Max \# of iterations of GA&$N_{it}$& 50\\
				\hline
				Environment parameters&$\eta_{LSO}, \eta_{NLSO}$ &1, 20&Environmental parameters& $\alpha, \beta$ &9.6, 0.28 \\
			    \hline
				\hline
			\end{tabular}
		\end{adjustbox}
		
		\label{sim_par}
	\end{table*}

\begin{figure*}[!h]
  \centering
  
  \subfloat[UAV and M-IRS placement at $t_1$]{\label{figur1}\includegraphics[width=.45\textwidth]{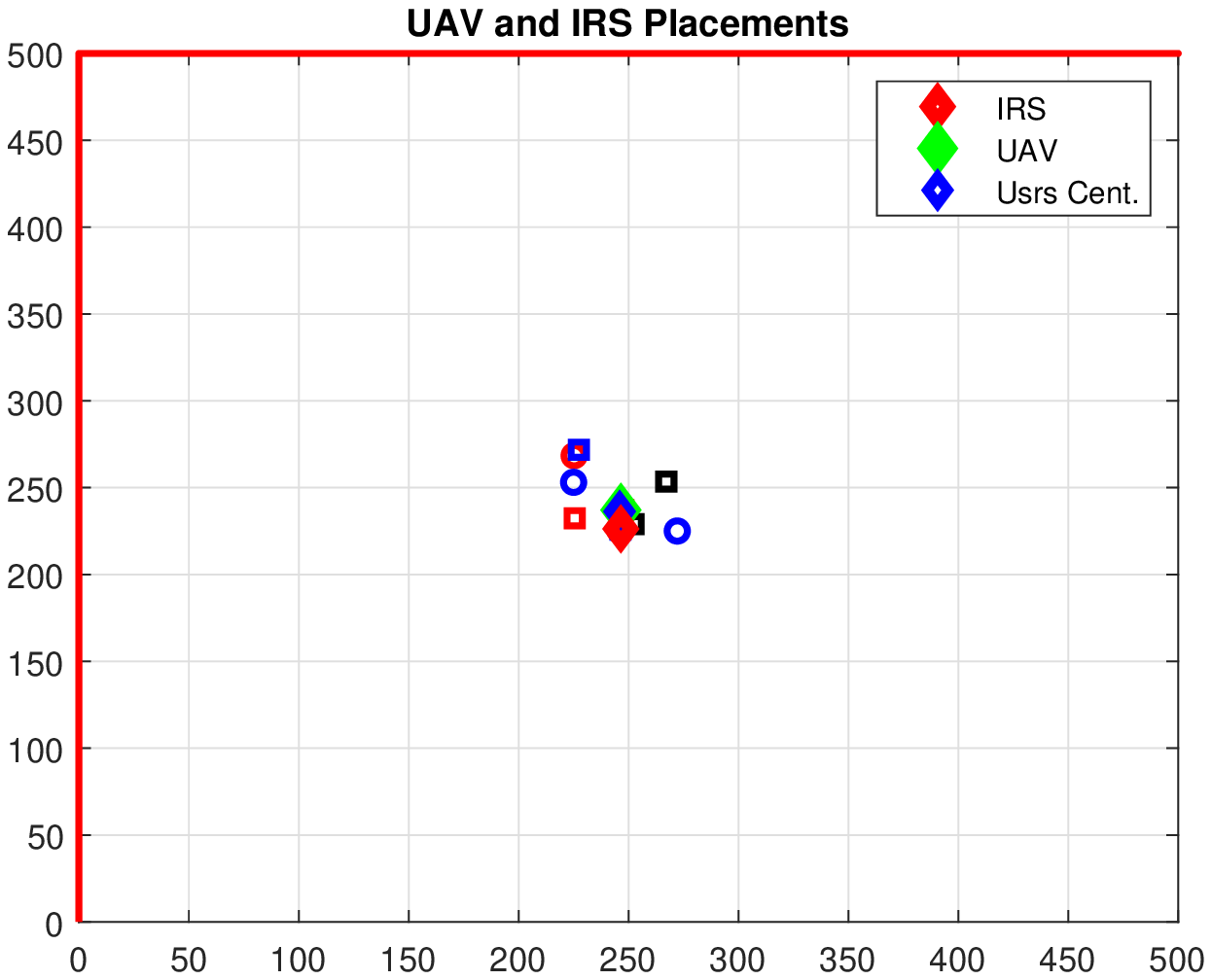}}
  \subfloat[UAV and M-IRS placement at $t_2$]{\label{figur2_2}\includegraphics[width=.45\textwidth]{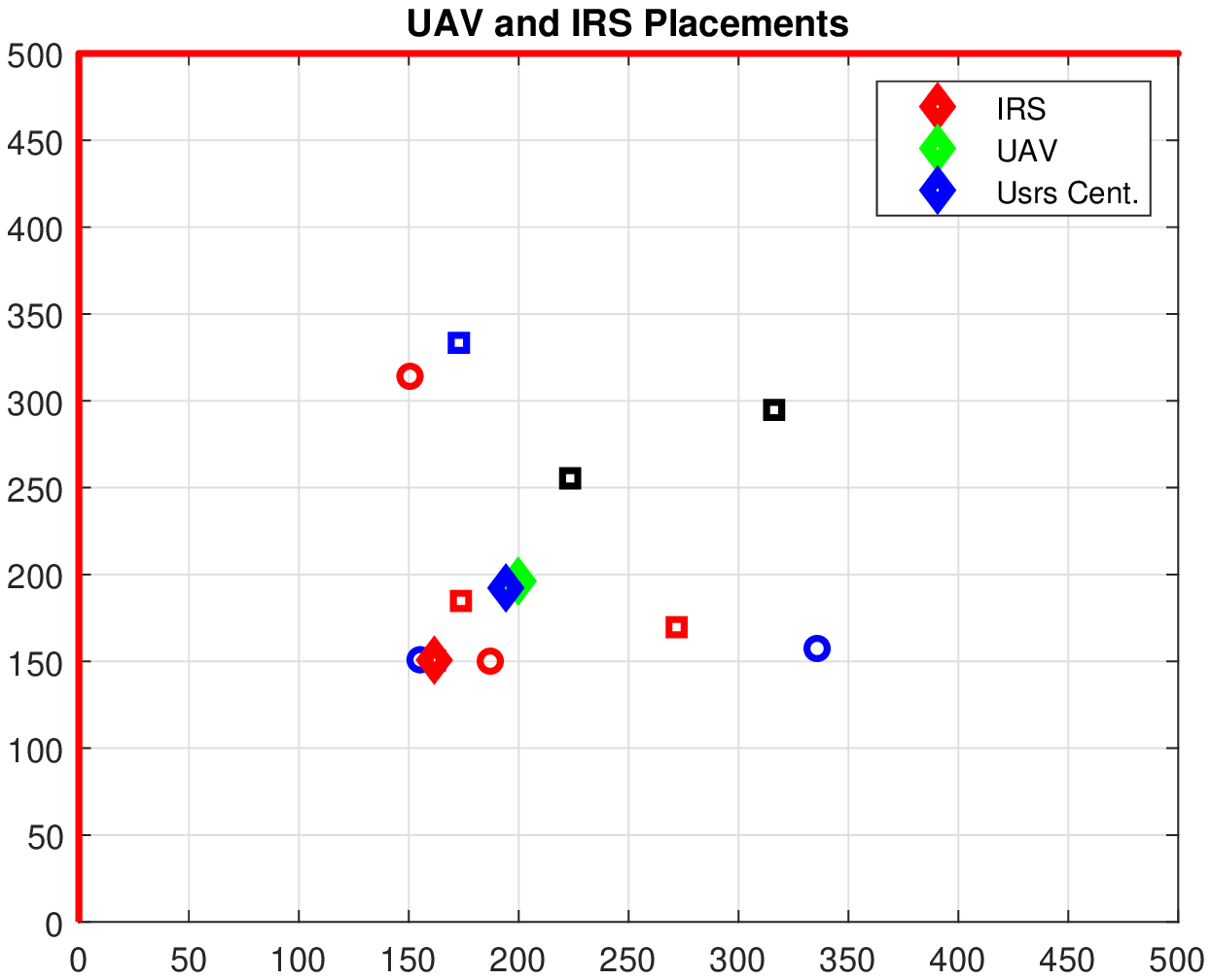}}
  \\
  \subfloat[UAV and M-IRS placement at $t_3$]{\label{figur3}\includegraphics[width=.45\textwidth]{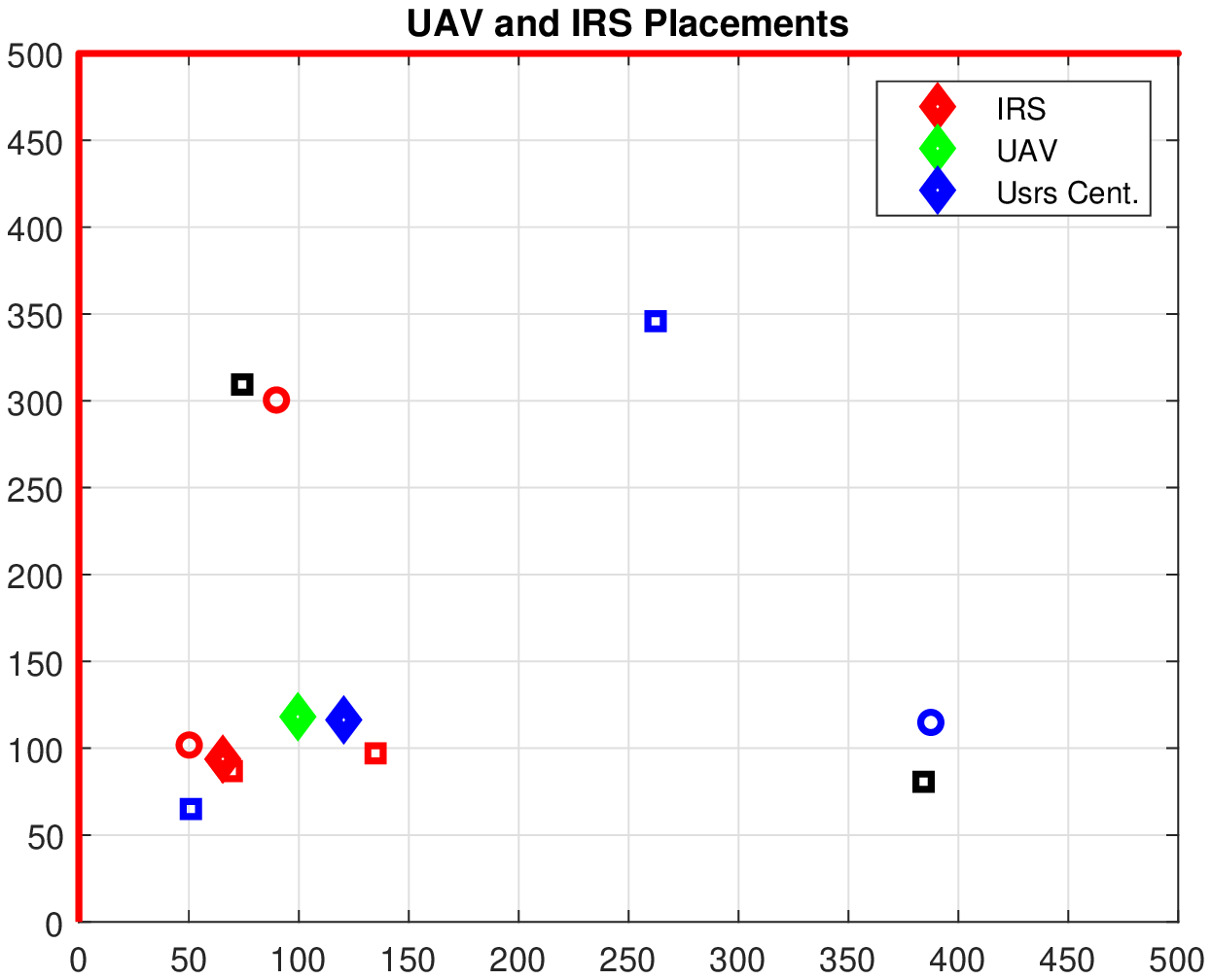}}
  \subfloat[UAV and M-IRS placement at $t_4$]{\label{figur4}\includegraphics[width=.45\textwidth]{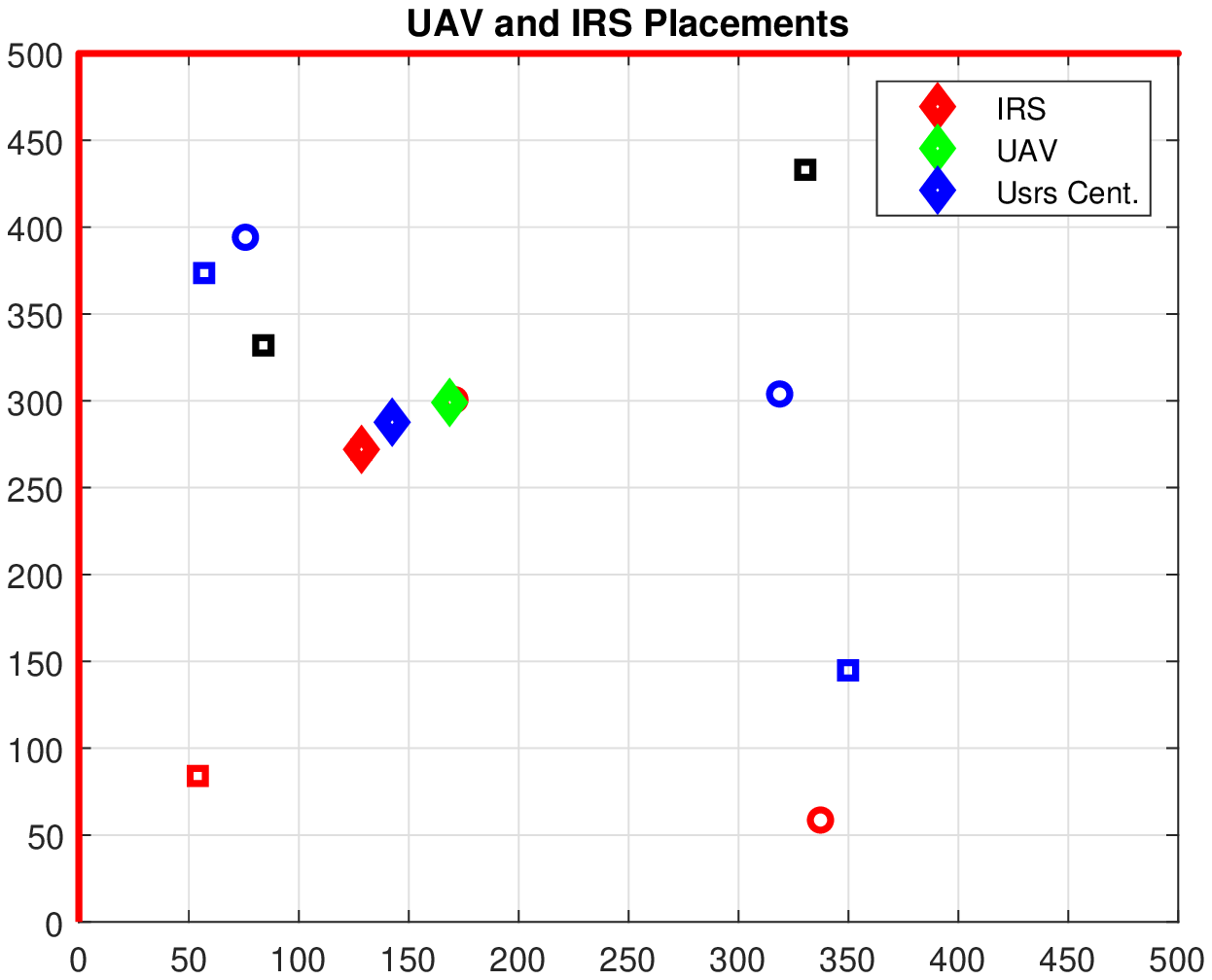}}
  \\
  \subfloat[UAV and M-IRS placement at $t_5$]{\label{figur5}\includegraphics[width=.45\textwidth]{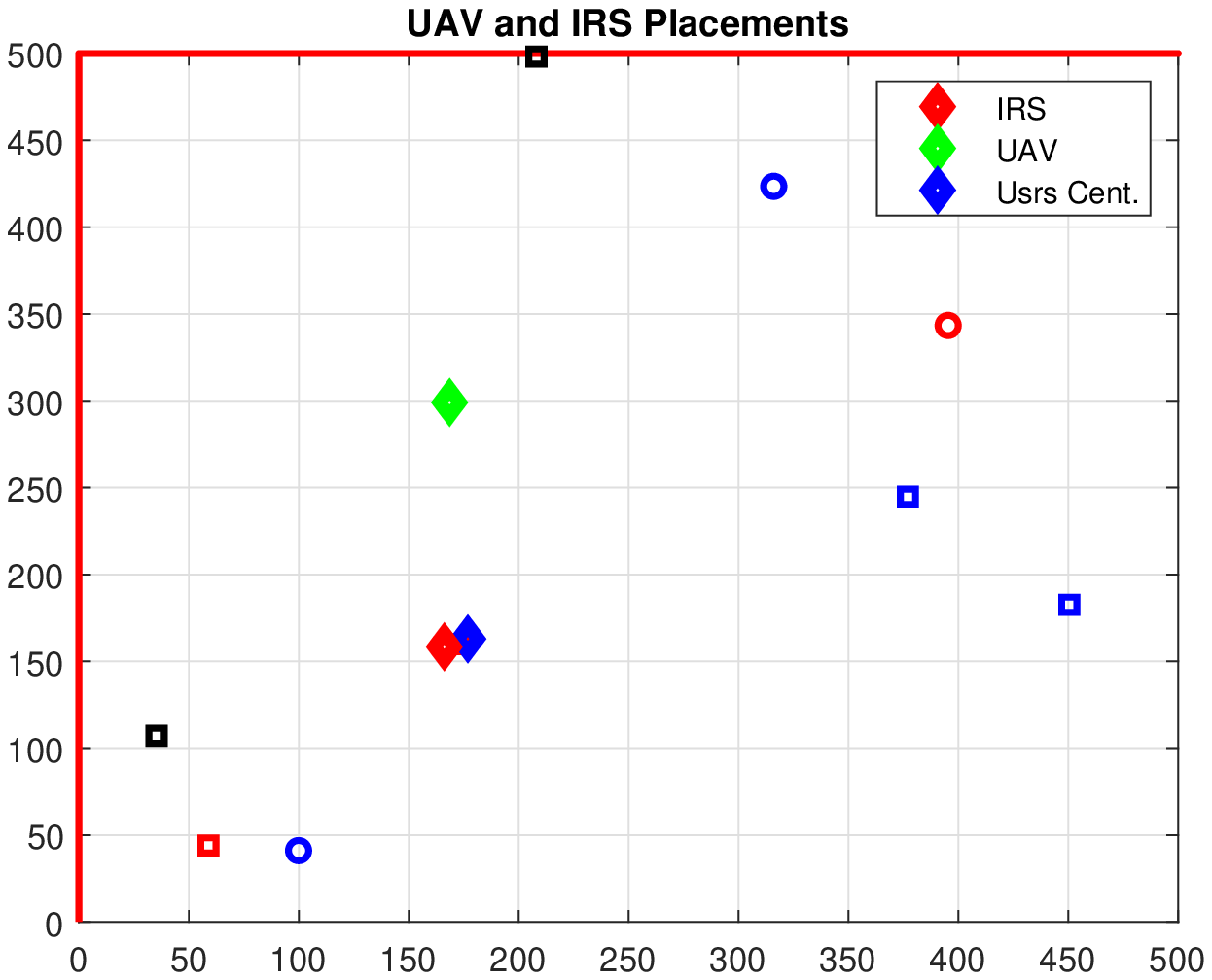}}
  \caption{UAV and M-IRS placement at $t_1$ to $t_5$ during users movement based  on  random  waypoint  mobility  model. The M-IRS placement is always near the largest gathering of users to maximize the total data rate between M-IRS and the users.  }
\label{UAV_IRS_P}
\end{figure*}

Figure~\ref{UAV_IRS_P} presents the UAV and M-IRS placements for times slot from $t1$ to $t5$. At each time slot, the GA is employed to evaluate the placement for UAV and IRS such that the average data rate for users is maximized. Specifically, at time slot $t1$, the users start their movement from the small subregion $50\times50$ as shown in Figure~\ref{usrs_distr}, the UAV and IRS locations at $t1$ is [246.5110 237.1115 100], and  [246.4811 226.2536 6], respectively. Moreover, the Figures~\ref{figur2_2} to \ref{figur5} show the users' locations during their mobility from time slot $t2$ to $t5$, respectively. For safety reasons, in this work , we set the minimum UAV height to 100m \cite{sawalmehijet,jasim2021survey}.

From Figure~\ref{UAV_IRS_P} we observed that the efficient M-IRS placement is always close to the largest gathering of users. This placement guarantees that the total data rate between M-IRS and the users is maximized.

\begin{figure*}[!h]
  \centering
  \subfloat[Top View UAV and M-IRS trajectory   ]{\label{figur1_traj}\includegraphics[width=0.45\textwidth]{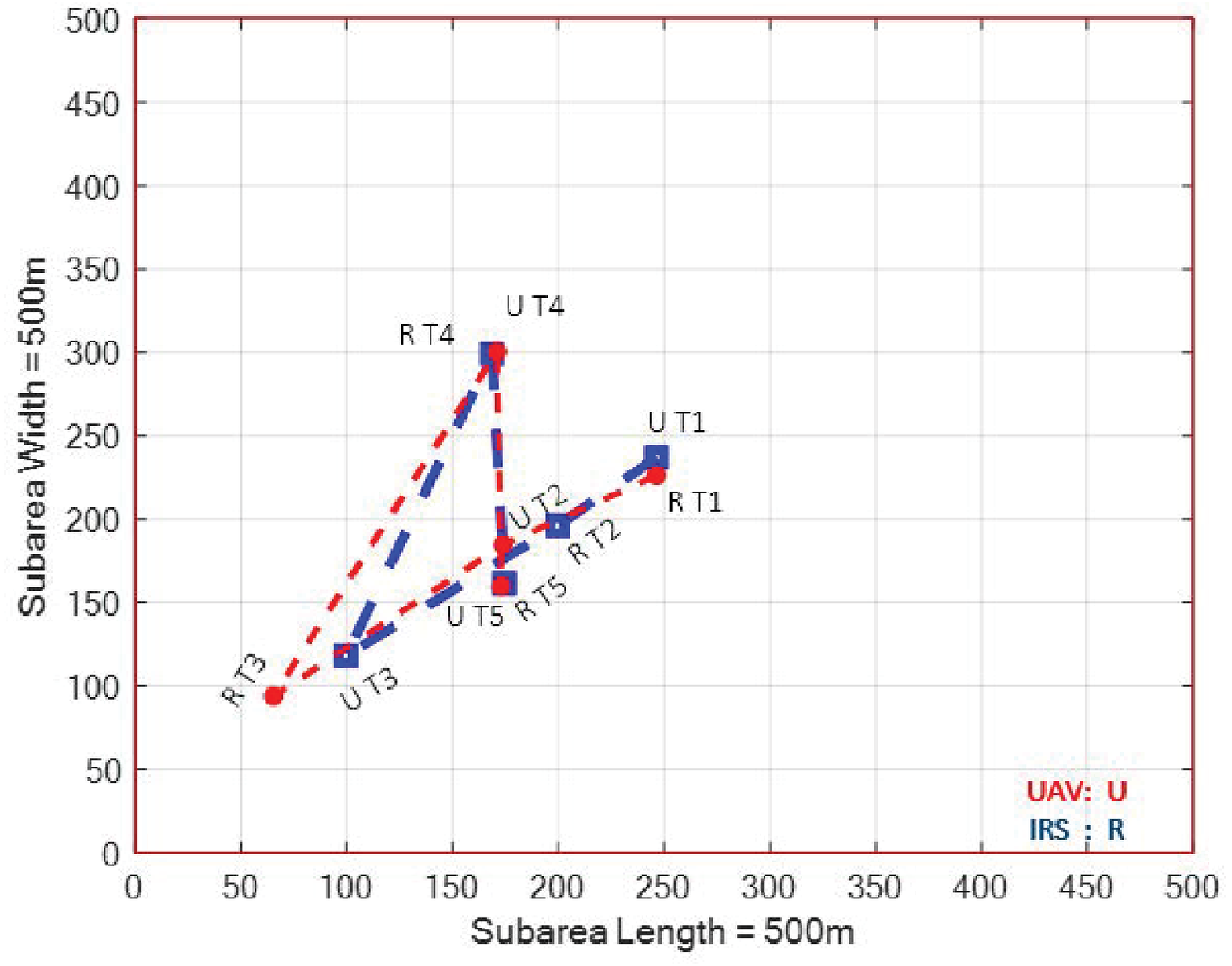}}
  \subfloat[3D UAV and M-IRS trajectory ]{\label{figur2_traj}\includegraphics[width=0.5\textwidth]{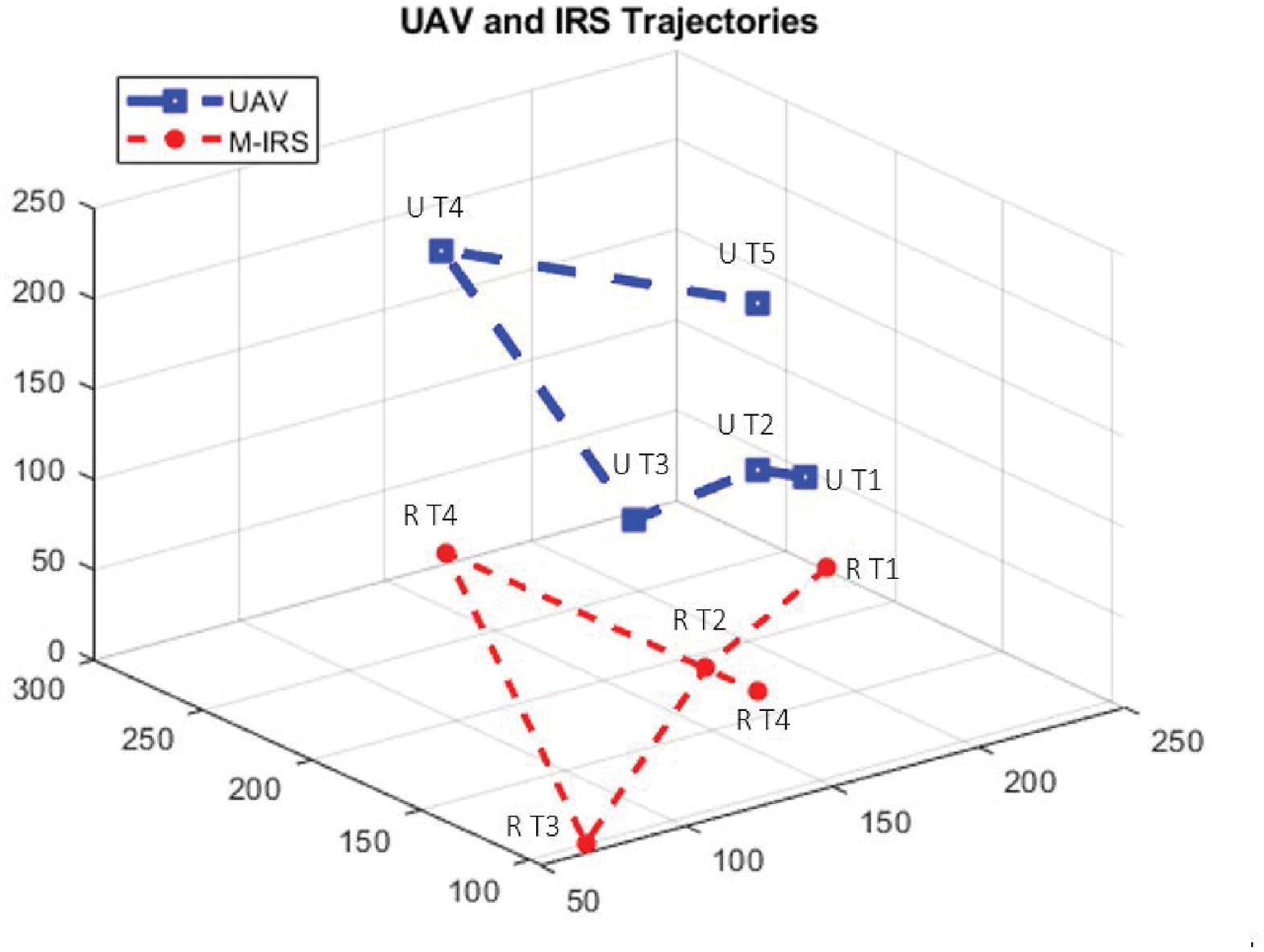}}
 \caption{UAV and M-IRS trajectory inside the coverage region from $t1$ to $t5$.}
\label{trajectory}
\end{figure*}

Figure~\ref{figur1_traj} presents a top view trajectory for UAV and M-IRS from $t1$ to $t5$. Moreover, Figure~\ref{figur2_traj} presents a 3D view trajectory for UAV and M-IRS from $t1$ to $t5$. In Figure~\ref{trajectory}, the dashed red and dashed blue colors present the trajectories for UAV and M-IRS respectively.

On the other hand, in this work, we evaluate and analyze the performance of the M-IRS-NOMA system. As a benchmarking, we compare the M-IRS-NOMA with the static IRS-NOMA (S-IRS-NOMA) and NOMA system without IRS in terms of average sum rate for each time slot. 

\begin{figure}[!h]
	\centering
	\includegraphics[scale=0.65]{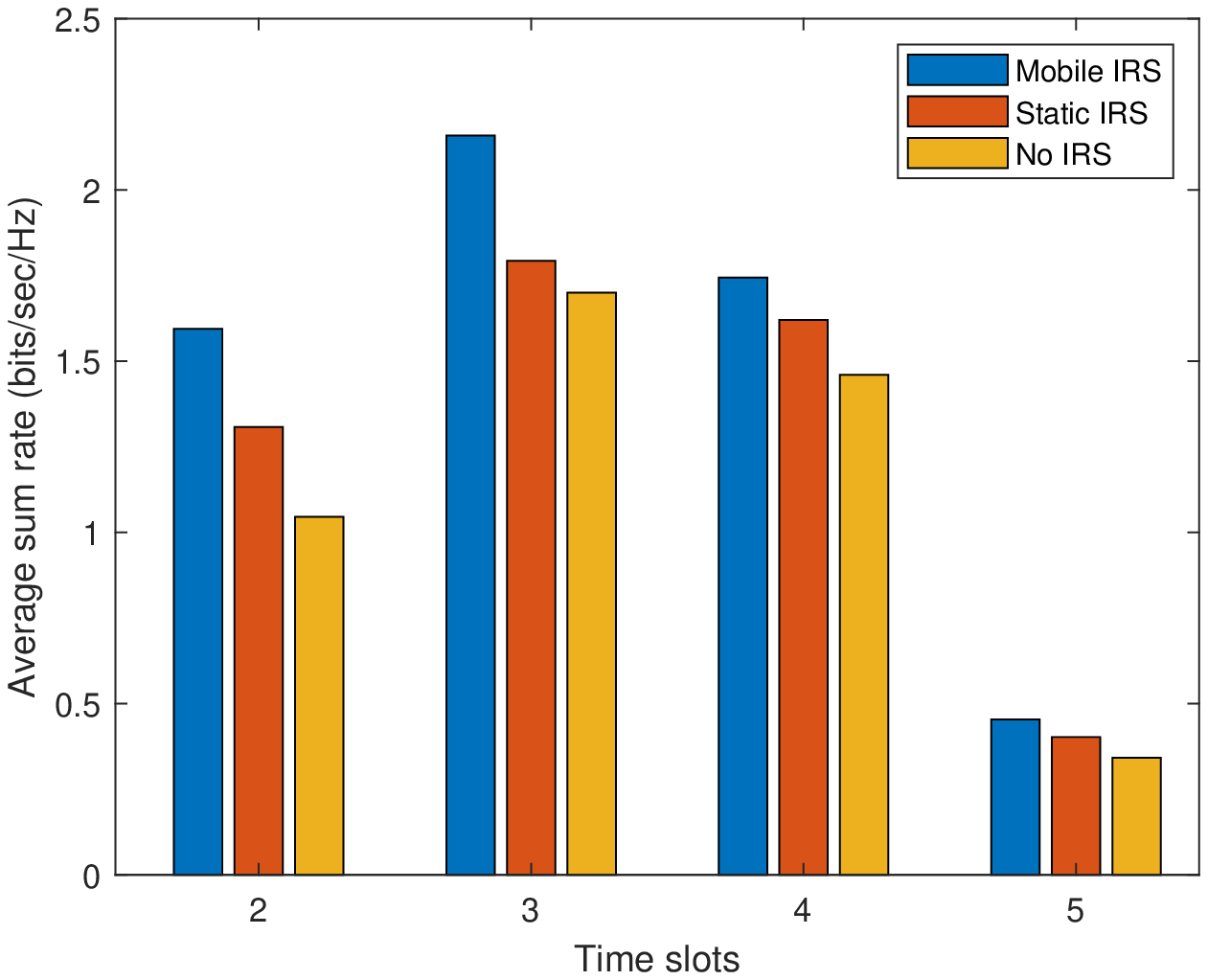}
	\caption{Average sum rate for different three scenarios mobile IRS, static IRS and no IRS. This Figure shows that the Mobile IRS NOMA outperforms the other two systems, static and no IRS scenarios. Mobile IRS NOMA can increase the average sum rate by $\sim$ $15\%$ on average compared with the static IRS. Also, the Mobile IRS NOMA outperforms the NOMA without the IRS system by $\sim$ $25\%$.}
	\label{Sum_rate}
	\end{figure}

For each scenario of the above-considered systems, the average sum rate is used to conduct a comparison among the three systems.  
Figure~\ref{Sum_rate} presents a bar graph to show the average sum rate of ground users for each time slot. It is clear from Figure~\ref{Sum_rate} that the M-IRS-NOMA has superior performance compared to the other two systems. The mobile IRS can increase the average sum rate by $\sim$ $15\%$ on average compared with the static IRS. Similarly, the M-IRS-NOMA outperforms the NOMA without IRS system by $\sim$ $25\%$ on average. The power fractions between the weak user and strong user in each pair in different time slots are shown in Figure~\ref{Power_fraction}. Table~\ref{time_uni} shows the average sum rate of the three considered systems. 

\begin{figure}[h]
	\centering
	\includegraphics[scale=0.65]{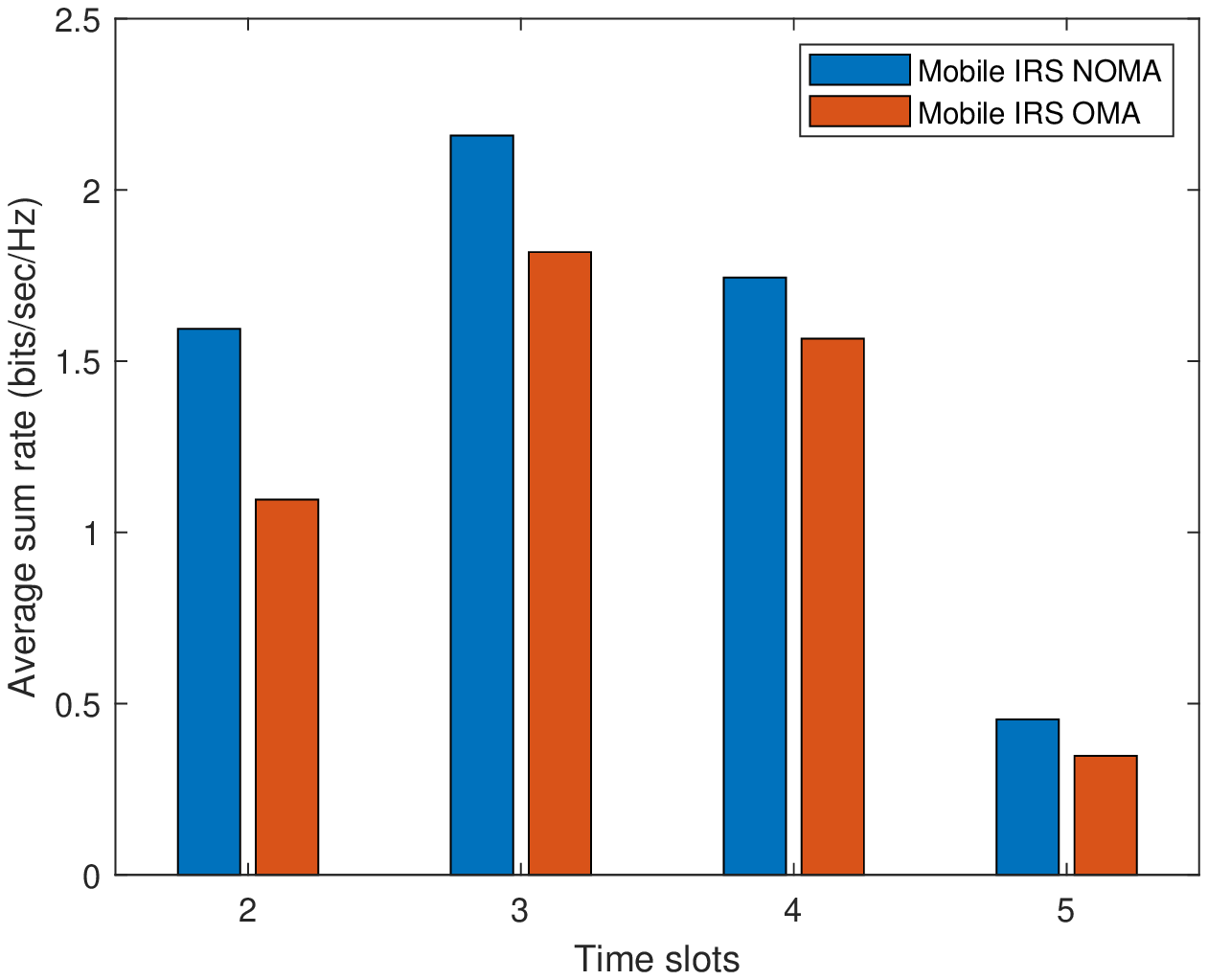}
	\caption{Comparison between Mobile-IRS-NOMA vs. Mobile-IRS-OMA. This Figure shows that the average sum rate for the Mobile-IRS-NOMA system is better than Mobile-IRS-OMA.  }
	\label{NOMA_vs_OMA}
	\end{figure}

\begin{figure*}[!h]
  \centering
  
  \subfloat[Power fraction for  $t_1$]{\label{figur1}\includegraphics[width=0.5\textwidth]{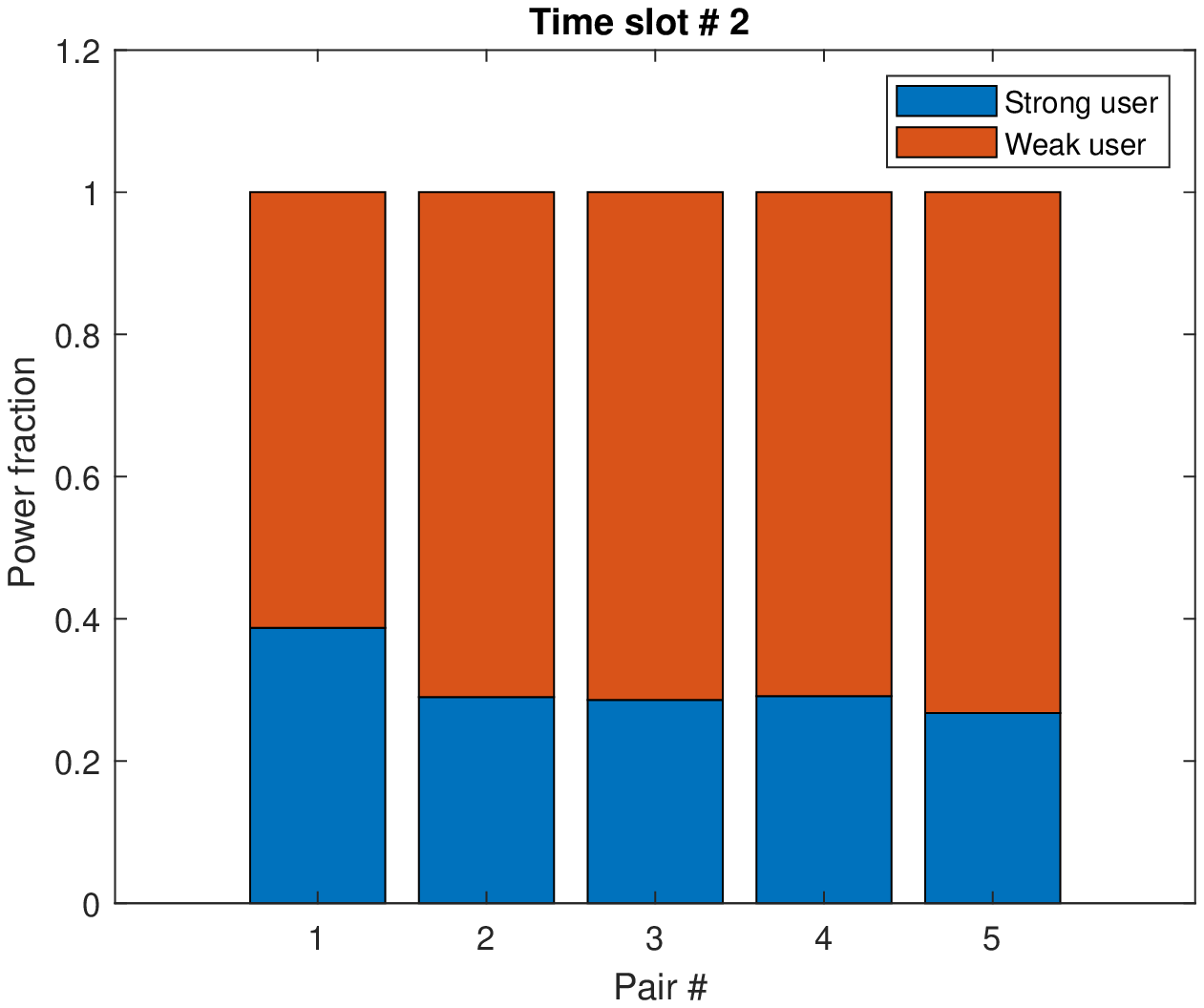}}
  \subfloat[Power fraction for  $t_2$]{\label{figur2}\includegraphics[width=0.5\textwidth]{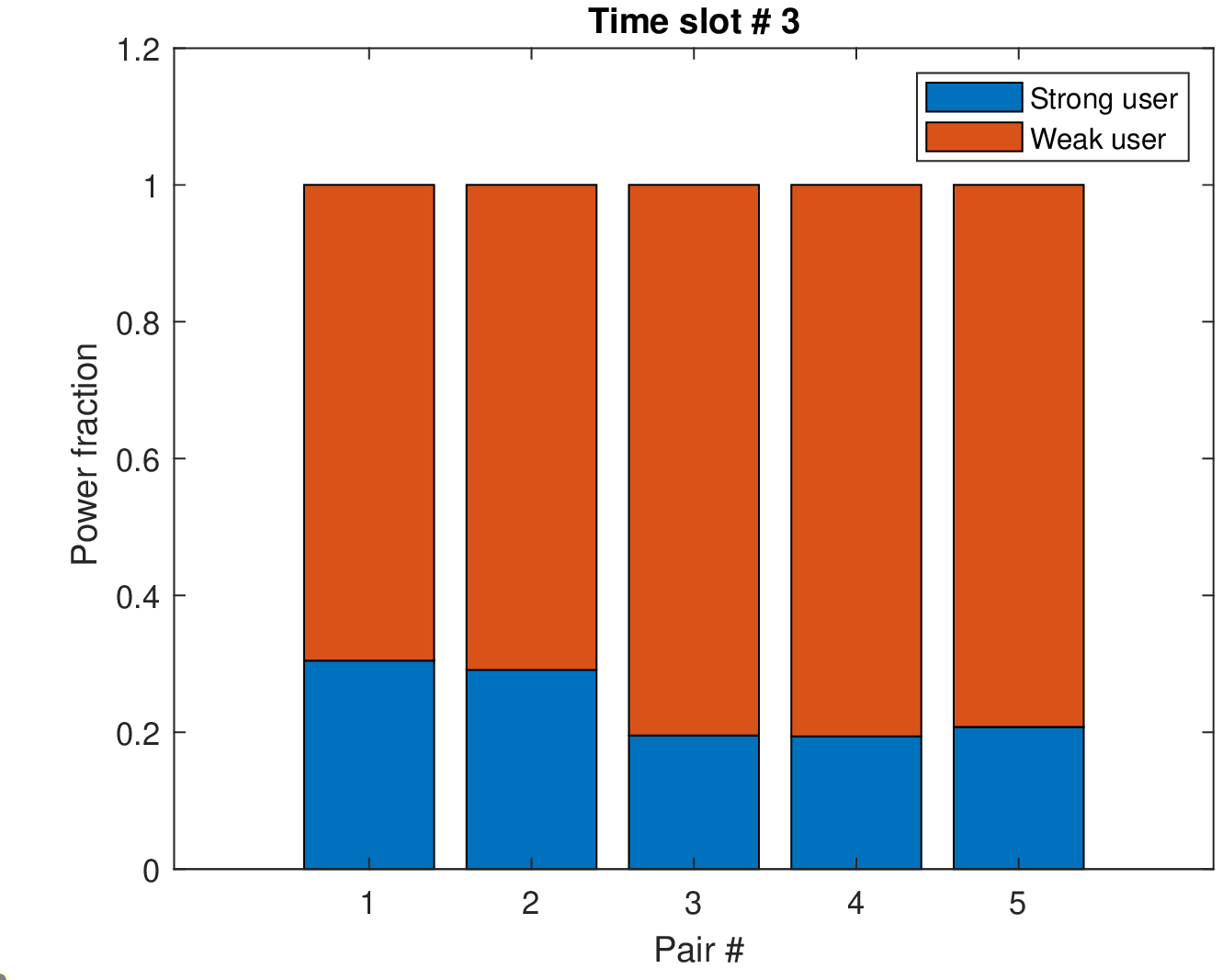}}
  \\
  \subfloat[Power fraction for  $t_3$]{\label{figur3}\includegraphics[width=0.5\textwidth]{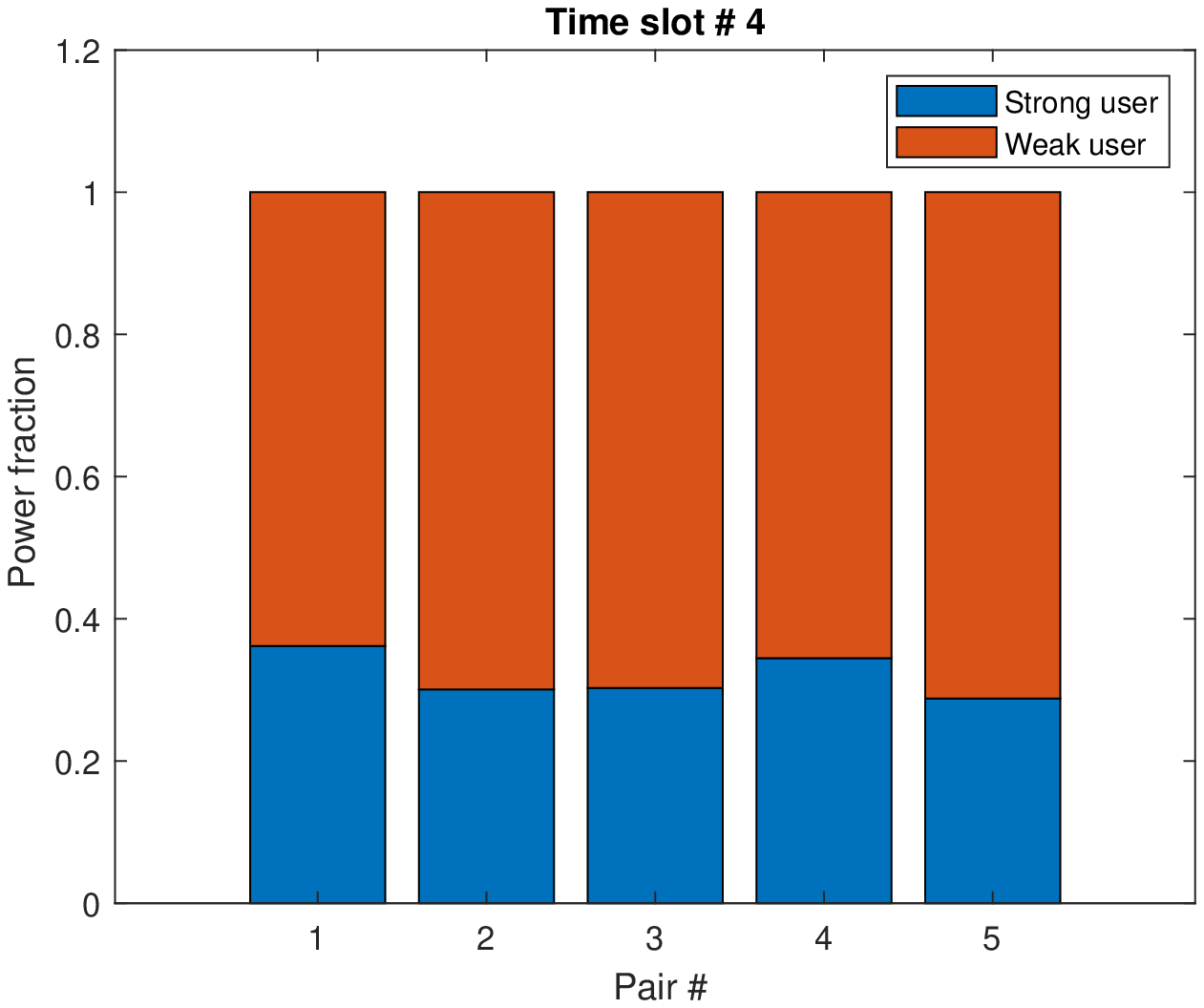}}
  \subfloat[Power fraction for $t_4$]{\label{figur4}\includegraphics[width=0.5\textwidth]{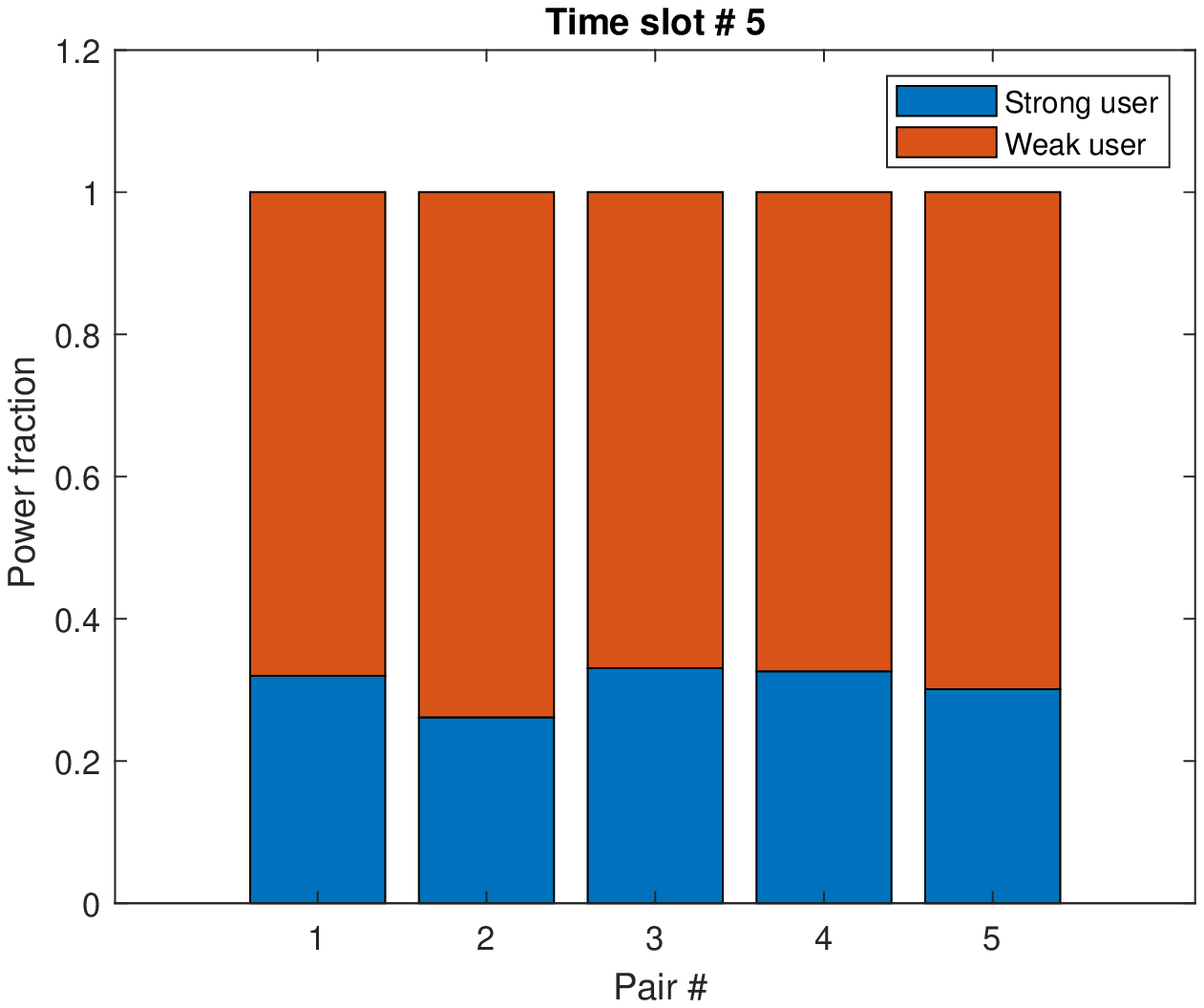}}
  \caption{The power fractions between the weak user and strong user for NOMA pairs from $t_1$ to $t_5$ during users movement based  on  random  waypoint  mobility  model. }
\label{Power_fraction}
\end{figure*}

Finally, it is well known that NOMA technology has better spectral efficiency than OMA, but this comes at the expense of the complexity of the receiver in the NOMA technique. On the other hand, the imperfect channel state information could degrade the performance of NOMA remarkably~\cite{reddy2021analytical}. For complete presentation of our problem, the performance of M-IRS-NOMA is compared with  M-IRS-OMA in terms of the average sum rate versus time slots as shown in Figure~\ref{NOMA_vs_OMA}. It can be observed from this Figure the M-IRS-NOMA outperforms the M-IRS-OMA.

\begin{table*}[!h]
    \renewcommand{\arraystretch}{1}
	\centering
	\label{time_uni}
	\caption{\MakeUppercase{Average sum data rate for different scenario}}
		\begin{adjustbox}{width=0.95\textwidth}
		\centering
		\tiny
		\begin{tabular}{|c|c|c|c|c|c|c|}
			
			\hline 
			&&\multicolumn{3}{c|}{{\textbf{Data Rate (bits/s/Hz)}}}&\multicolumn{2}{c|}{{\textbf{Percentage}}}\\
			\textbf{S}& \textbf{Time Slot}&\textbf{M-IRS} & {\textbf{S-IRS}}  & {\textbf{No-IRS}}&\textbf{M-IRS Vs. S-IRS} & {\textbf{M-IRS Vs. NO-IRS}}  \\
			\hline
			\hline
			1& $t1$ & 1.6 & 1.3 &  1  & 18.75\%  & 35\% \\
			\hline
			2& $t2$ & 2.2 & 1.8 & 1.7 & 16.67\% & 21.3\% \\
			\hline
			3& $t3$ & 1.75& 1.6 & 1.46& 9.37\%  & 18.12\%\\
			\hline
			4& $t4$ &0.455& 0.4 & 0.34& 13.5\%  & 25.11\%\\
			\hline
		\end{tabular}
	\end{adjustbox}
\end{table*}

\section{LESSONS LEARNED}
\label{lessons}
In this section, the key findings and the lessons learned are highlighted as listed below:
\begin{itemize}
    \item IRS mounted on mobile vehicle: There are several challenges for mounted IRS on UAVs, such as; UAV payload limitations; moreover, the UAV and IRS could deviate in strong wind conditions. Therefore, IRS mounted on a mobile ground vehicle Mobile-IRS could be considered a promising solution to tackle these challenges and improve the overall network throughput. Specifically, Mobile-IRS outperforms the static IRS in terms of network average sum rate. 
    
    \item NOMA with a users-pairing scheme is considered one of the significant future trends for wireless networks 5G and beyond to fulfill the demands of these networks in the following factors; 1) spectral efficiency, 2) low latency, 3) user fairness, and 4) high bandwidth, to maximize the overall network data rate. NOMA can attain more reasonable fairness and better spectral efficiency compared to OMA networks.

    \item Most of the previous studies considered static user scenarios; this assumption limits the applicability of the mobile users model in the real scenarios, where the users are generally mobile and use wireless devices during their movement. Therefore, future research studies need to consider the mobility of users and use the appropriate mobility model for them. However, researchers can use artificial intelligence methods to predict and generate a real scenario that represents users’ mobility as an open research direction.

    
    %
    
    \item Using heuristic approaches such as the Genetic  Algorithm (GA) to find an efficient UAV trajectory and an efficient path for the M-IRS  will reduce the computation complexity and, thus, significantly reduces execution time compared with the optimal methods used to find the trajectories for UAV and M-IRS in providing wireless connectivity for mobile users during their movement.

\end{itemize}

\section{Conclusion}
\label{sec7}
Unlike previous studies, this paper studies the utilization of a single UAV and an M-IRS in an IoT-6G wireless network. The optimization problem aims to find an efficient trajectory for the UAV, an efficient path for the M-IRS, and users’ power allocation coefficients that maximize the average data rate for mobile ground users. To represent the users’ movements inside the coverage area, we employ the individual movement model (Random Waypoint Model). We also show that the dynamic power allocation technique outperforms the fixed power allocation
technique in terms of network average sum rate. To find an efficient trajectory for the UAV and an efficient path
for the M-IRS during providing wireless connectivity for mobile users, we propose an efficient approach using a Genetic Algorithm. The proposed methodology for providing wireless coverage using an M-IRS assisted UAV system is expected to use in smart cities.


\bibliographystyle{IEEEtran}
\bibliography{ref}


\end{document}